# Dual-wavelength pumped highly birefringent microstructured silica fiber for widely tunable soliton self-frequency shift

Olga Szewczyk, Piotr Pala, Karol Tarnowski, Jacek Olszewski, Francisco Senna Vieira, Chuang Lu, Aleksandra Foltynowicz, Paweł Mergo, Jarosław Sotor, Grzegorz Soboń, and Tadeusz Martynkien

*Abstract*— We report the design of a microstructured silica-based fiber for widely tunable soliton self-frequency shift, suitable for pumping with two most common fiber laser wavelengths: 1.04 µm and 1.55 µm. Depending on the pump source, the output spectrum can be continuously tuned up to 1.67 µm (pump at 1.04 µm) or 1.95 µm (pump at 1.55 µm) in the same 1.5 m-long fiber sample, with pump-to-soliton conversion efficiency higher than 20%. The fiber is highly birefringent, which results in an excellent polarization extinction ratio of the soliton, reaching 26 dB. The shifted solitons have a high degree of coherence confirmed by pulse-to-pulse interference measurement. The available soliton tuning range covers the wavelengths inaccessible for fiber lasers, e.g., 1.3 µm and 1.7 µm, highly important for multi-photon microscopy and imaging. Our work shows that it is possible to design and fabricate one universal optical fiber that supports soliton shift when pumped at two different wavelengths separated by over 500 nm.

*Index Terms*—Fiber nonlinear optics, optical solitons, optical wavelength conversion.

## I. Introduction

THE soliton self-frequency shift (SSFS), experimentally demonstrated in 1986 [1], is a nonlinear phenomenon in dispersion-engineered optical fibers. The SSFS results from stimulated Raman scattering and occurs when an optical fiber is pumped with ultrashort laser pulses. Due to the Raman gain in the fiber, the short-wavelength part of the soliton spectrum pumps the long-wavelength part, resulting in an increasing redshift of the soliton spectrum. Since the Raman gain in silica has a very broad bandwidth, the effect occurs already at sub-picosecond excitation. The SSFS enables continuous wavelength tuning of ultrashort laser pulses with high efficiency (up to 97% [2]) in a broad spectral range with preserved nearly-transform-limited duration. The importance of this phenomenon for laser technology has been confirmed in many applications, like three-photon microscopy for deep-tissue imaging [3] or mid-infrared optical frequency comb generation [4]. The SSFS enables to obtain non-standard wavelengths located outside the gain bandwidths of existing laser media (i.e., Yb, Er, Tm, Ho), e.g., 1.1–1.4 µm, 1.6–1.7 µm or beyond 2.1 µm. Nishizawa and Goto demonstrated pumping of a polarization-maintaining (PM) optical fiber with a 1.55 µm source to obtain the soliton shift up to 2 µm [5]. Since then, several realizations of the SSFS effect have been demonstrated, aiming at higher output powers, shorter pulses, and broader tuning range. Nguyen et al. presented the SSFS effect in a large mode area (LMA) fiber resulting in tunable solitons covering the range of 1.6–1.7 µm with pulse energies of 13 nJ and 550 mW average power [6]. Tan et al. generated SSFS in a nonlinear fiber pumped by an Er-doped fiber laser (EDFL) and amplified the output pulses at 1970 nm achieving 1.02 µJ pulse energy with a pulse duration of 241 fs and 35.4 W average power [7]. The spectral range beyond 2 µm was also reached in numerous setups utilizing: a cascade of silica and germanium optical fibers [8], a suspended core fiber pumped by Tm and Er sources [9], a microstructured tellurite fiber together with the hybrid Er/Tm laser system [10] or fluoride fibers achieving the shift even up to 4.3 µm [11]. Simultaneously there has been growing interest in pumping SSFS fibers by Yb-doped fiber lasers (YDFLs). Lim et al. reported a system with photonic crystal fiber pumped with a femtosecond Yb oscillator [12]. Later, Dawlaty et al. [13], as well as Takayanagi et al. [14], presented widely tunable solitons in the range of 1.0–1.7 µm. The duration of the output pulses was in the 100–150 fs range. However, both these setups employed non-polarization maintaining fibers.

The majority of research work concerning the SSFS effect focuses on scaling up the power and shortening the duration of the output pulses. Yet, high-quality linear polarization and high

This work was supported in part by the National Centre for Research and Development (POIR.04.01.01-00-0037/17) and partially by the Knut and Alice Wallenberg Foundation (KAW 2015.0159).

Olga Szewczyk, Jarosław Sotor, and Grzegorz Soboń are with the Laser & Fiber Electronics Group, Wrocław University of Science and Technology, 50-370 Wrocław, Poland (e-mail: olga.szewczyk@pwr.edu.pl; jaroslaw.sotor@pwr.edu.pl; grzegorz.sobon@pwr.edu.pl).

Piotr Pala, Karol Tarnowski, Jacek Olszewski, and Tadeusz Martynkien are with the Faculty of Fundamental Problems of Technology, Wrocław University of Science and Technology, 50-370 Wrocław, Poland (e-mail: piotr.pala@pwr.edu.pl; karol.tarnowski@pwr.edu.pl; jacek.olszewski@pwr.edu.pl; tadeusz.martynkien@pwr.edu.pl).

Francisco Senna Vieira, Chuang Lu, and Aleksandra Foltynowicz are with the Department of Physics, Umeå University, 901 87 Umeå, Sweden (e-mail: aleksandra.foltynowicz@umu.se; chuang.lu@umu.se; francisco.sennavieira@vtt.fi).

Paweł Mergo is with the Laboratory of Optical Fiber Technology, Maria Curie-Skłodowska University, 20-031 Lublin, Poland (pawel.mergo@poczta.umcs.lublin.pl).

degree of coherence are also key parameters of a laser source for real-life applications. The design flexibility of microstructured silica fibers allows for improving the tuning range of the shifted solitons in contrast to conventional fibers. Moreover, adequately designed microstructured fibers can be operated at different pump wavelengths. Numerous experiments use pump sources operating at two common optical wavelengths, 1.06 or 1.55 μm, to generate solitons in optical fibers [12–18]. However, to the best of our knowledge, none of them have presented a fiber suitable for pumping at both these spectral ranges. In this paper, we report the development and fabrication of a highly birefringent microstructured silica fiber for SSFS designed for pumping with light sources operating at 1 μm and 1.55 μm. The solitons are highly coherent and can be continuously tuned up to 1.7 μm (pump at 1.04 μm) or 1.95 μm (pump at 1.55 μm) in the same, 1.5-m long fiber. The wavelength can be further shifted up to 2.07 μm by using a longer fiber sample.

## II. Fiber Design and Characterization

### A. Design of the SSFS Fiber

Silica-based microstructured optical fibers (MOFs) allow circumventing limits arising from the dispersion properties of conventional fibers. In particular, the SSFS effect requires anomalous dispersion, which occurs only above 1.3 μm in standard optical fibers. The MOF technology enables flexible dispersion tailoring and shifting the zero-dispersion wavelength (ZDW) to much shorter wavelengths, which allows the use of the popular Yb-doped pump lasers for SSFS generation. It is done by adequately selecting the fiber constructional parameters: lattice pitch ($\Lambda$) and air filling factor ($ff = d/\Lambda$, where d denotes the cladding air holes diameter). Considering the requirement of anomalous dispersion for pumps at 1.04 μm and 1.55 μm, we aimed at obtaining structure designs that fulfilled the following criteria:

i. chromatic dispersion: ZDW below 1 μm and maximized chromatic dispersion in the entire soliton tuning range (1–2 μm);
ii. a phase modal birefringence larger than $2 \cdot 10^{-4}$ at the pump wavelength to minimize coupling between the polarization modes;
iii. a low transmission loss in the whole spectral range of interest (below 0.1 dB/m at 2.0 μm);
iv. relatively simple cross-section features to allow effective and repeatable manufacturing of the fiber.

To design the optimal fiber structure, we performed numerical simulations utilizing a fully vectorial approach. We employed the finite element method (FEM) based on edge/nodal hybrid elements and anisotropic perfectly matched layers boundary conditions. During these simulations, the spectral dependence of the refractive index of pure silica glass was taken into account. First, we found the propagation constants and field profiles of the fundamental polarization modes. Next, we calculated the modal birefringence and chromatic dispersion profiles. We started the design process with the analysis of the non-birefringent structures. We calculated chromatic dispersion profiles and effective nonlinear coefficient ($\gamma$) with the air filling factor in the 0.4–0.9 range for the different pitch values in the range of 1.5–3.1 μm. High air filling factors are needed to assure high waveguide contribution to the chromatic dispersion and fulfill the target requirements.

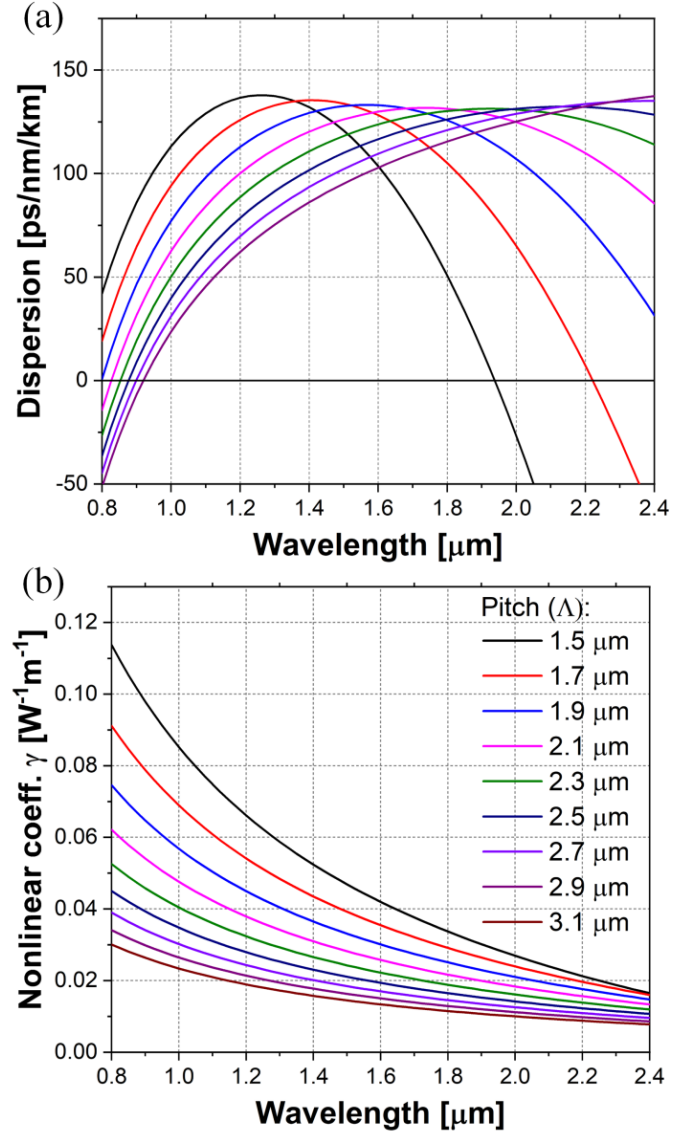

Fig. 1 Calculated properties of MOF: (a) chromatic dispersion, (b) nonlinear coefficient, for different pitch value $\Lambda$ and filling factor fixed to $ff = 0.8$. The color legend from (b) also applies to (a).

As an example, Fig. 1 shows the calculation results of chromatic dispersion (a) and the nonlinear coefficient (b) for different pitch values and fixed air-filling factor $ff = 0.8$. With increasing pitch, we observe decreasing chromatic dispersion at 1.06 μm and decreasing effective nonlinearity coefficient in the considered wavelength range. As a consequence, soliton tuning using a 1.06 μm pump would be the most effective in a MOF with $\Lambda = 1.5$ μm. However, in this structure, a second ZDW appears around 1.9 μm, limiting the soliton tuning range. Moreover, to allow simultaneous soliton tuning from 1.55 μm, it is required to keep high chromatic dispersion also at this wavelength. Finally, our previous experience with soliton generation in small-core MOFs showed that $\Lambda > 2.5$ μm is necessary to get high coupling efficiency to the fundamental mode of the fiber [18]. This can be intuitively explained by recalling that the core diameter is approximately $d_{core} = 2\Lambda-d$. Consequently, we targeted the fiber with $\Lambda$ = 2.5–2.7 μm, $ff$ = 0.8, and $d_{core}$ about 3 μm. Due to the relatively high air filling factor of the whole photonic structure, it was impossible to induce birefringence by enlarging holes adjacent symmetrically to the core [19]. Therefore, the high birefringence was introduced by diminishing the diameter of two cladding holes symmetrically adjacent to the fiber core. This can be expressed with decreased filling factor $ff_s = d_s/\Lambda$. The performed phase modal birefringence simulations of the structures with small holes with $ff_s$ = 0.5–0.6 indicate that modal birefringence exceeds $3\cdot10^{-4}$ at 1.06 μm and increases with wavelength, which is typical for highly birefringent MOFs. The final fiber design is presented in Fig. 2(a). The five rings of air holes surrounding the fiber core were introduced to ensure the reliability of the MOF structure during the fabrication process. The holes in the outer ring tend to collapse, decreasing the air-filling factor. Therefore, five rings are needed to keep desirable geometrical parameters. Moreover, the five air-holes rings guarantee that the confinement loss for the fundamental polarization modes does not exceed 0.1 dB/m at 2.0 μm.

Finally, the designed MOF has the following geometrical parameters: the pitch distance $\Lambda$ = 2.70 μm, the small holes' diameter $d_s$ = 1.44 μm ($ff_s$ = 0.53), and the diameter of remaining cladding holes $d$ = 2.2 μm ($ff$ = 0.81). The design of the structure is depicted in Fig. 2(a). The fiber was fabricated at the Laboratory of Optical Fibers Technology, University of Maria Curie-Sklodowska in Lublin, Poland, via the standard stack and draw method. We used a synthetic silica glass type VI (Heraeus F300ES). The photonic structure has been produced with suitably prepared capillaries with an outer diameter of 1 mm and inner diameters 0.78 mm and 0.65 mm for big and small holes, respectively. The thermal bonding of all capillaries and overcladding processes were performed on a drawing tower. The scanning electron microscope (SEM) image of the fiber's cross-section is depicted in Fig. 2(b), showing excellent agreement between the fabricated and designed structure.

*B. Fiber characterization*

The transmission loss spectrum of the fabricated fiber (Fig. 3) was measured by the cut-back method using a broadband supercontinuum source (NKT Photonics EXR-20) and an optical spectrum analyzer (Yokogawa AQ6375, OSA). The transmission loss is lower than 0.1 dB/m in the range of interest (1200 – 2000 nm), except at the wavelengths corresponding to the OH absorption bands. The spectral attenuation loss can be further reduced by employing a core rod made from glass with significantly lower OH- content, e.g. Heraeus Suprasil F500 (0.02 ppm compared to 0.2 ppm for F300ES [20]).

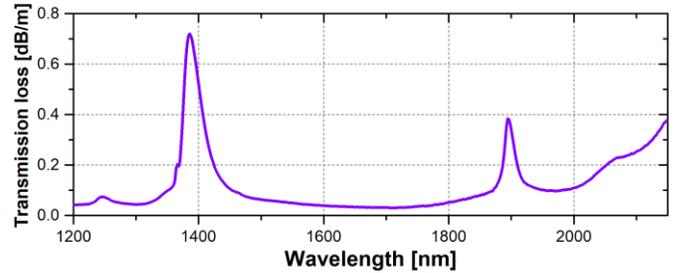

Fig. 3. Transmission loss of the manufactured fiber.

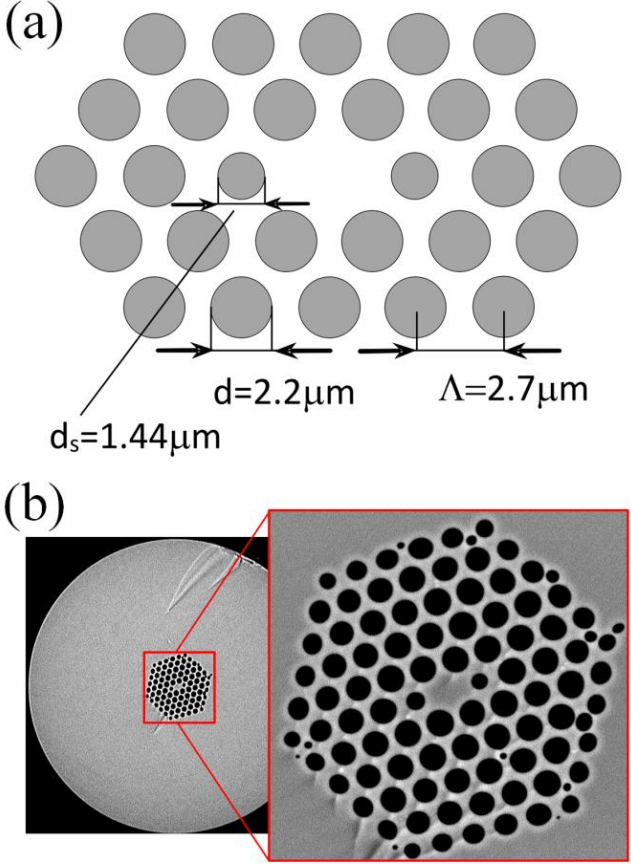

Fig. 2. (a) Constructional parameters of the MOF: lattice pitch ($\Lambda$), $d$ – dimeter of cladding holes, $d_s$ - diameter of small holes adjacent symmetrically to the core; (b) Scanning electron microscope image of the fiber cross-section.

To experimentally characterize the manufactured fiber, we measured the spectral dependence of chromatic dispersion for both polarization modes. The dispersion measurements were conducted by the white light interferometric technique in the 0.85–2.10 μm spectral range. We used a dispersion-balanced Mach–Zehnder interferometer (MZI) with an adjustable reference arm, while the investigated fiber of known length $L$ (typically 0.8–1.2 m long) was placed in the other arm of the MZI [21]. We acquired a series of spectral interferograms using two spectrometers (Ocean Insight NIRQuest 512 and Avesta ASP IR) to estimate the equalization wavelength $\lambda_{eq}$ as a function of the path length difference $\Delta L(\lambda)$. Afterward, we fit the obtained data $\Delta L(\lambda_{eq})$ with a five-term power series to calculate the dependence of the chromatic dispersion vs. wavelength, using the relation $D = (L\cdot c)^{-1}\cdot(d\Delta L/d\lambda)$, where $c$ denotes the velocity of light. Figure 4 presents the measured and calculated spectral dependence of dispersion for both polarization modes ($LP_{01}^X$, $LP_{01}^Y$). The ZDW equals 932 nm and 945 nm for the $LP_{01}^X$ and $LP_{01}^Y$ modes, respectively. The two arrows (blue and green) indicate the pump wavelengths at 1040 and 1560 nm, respectively.

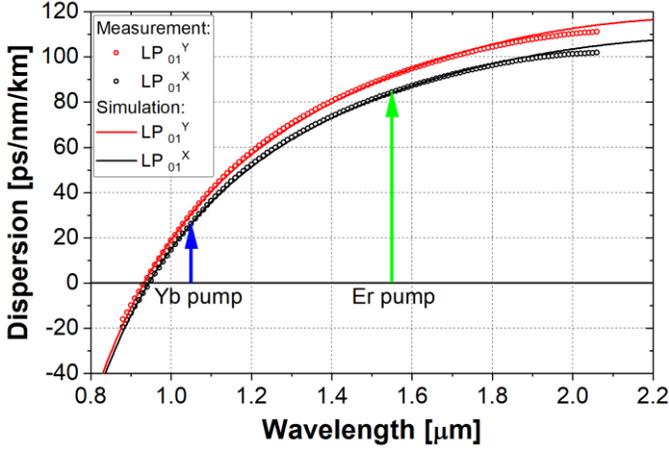

Fig. 4. Measured (dotted curve) and calculated (solid curve) dispersion profiles of the fabricated SSFS fiber for two orthogonal polarization modes, with indicated pumping wavelengths.

Once the spectral dependence of $\Delta L_x(\lambda)$ and $\Delta L_y(\lambda)$ was known, we estimated the group modal birefringence as a function of wavelength using the relation $G(\lambda) = (\Delta L_x(\lambda) - \Delta L_y(\lambda))/L$. The phase modal birefringence $B(\lambda)$ was measured using the lateral force method in the spectral domain [22]. In this method, applying a point-like force induces significant energy coupling from the initially excited mode to the orthogonally polarized mode, which creates an interference pattern after passing the analyzer. In response to the displacement of the coupling point along the tested fiber by the distance $dL$, spectral fringes shift is observed by $dM$ fringes (observed at a particular wavelength). Hence the beat length for respective spatial mode at specific wavelength can be determined from the relation $L_B = dL/dM$ and B can be calculated form relation $B = \lambda/L_B$. Since the group modal birefringence $G(\lambda)$ is related to the phase modal birefringence $B(\lambda)$ via the expression $G(\lambda) = -\lambda^2 d[B(\lambda)/\lambda]/d\lambda$ [23], we can obtain the relative wavelength dependence of the phase modal birefringence. Furthermore, the known value of phase modal birefringence at a specific wavelength can be used for absolute calibration and obtaining the wavelength dependence of the phase modal birefringence $B(\lambda)$ [23]. The wavelength dependences of phase and group modal birefringence are shown in Fig. 5(a). Owing to a relatively high modal birefringence, we did not observe polarization cross talk in the investigated spectral range.

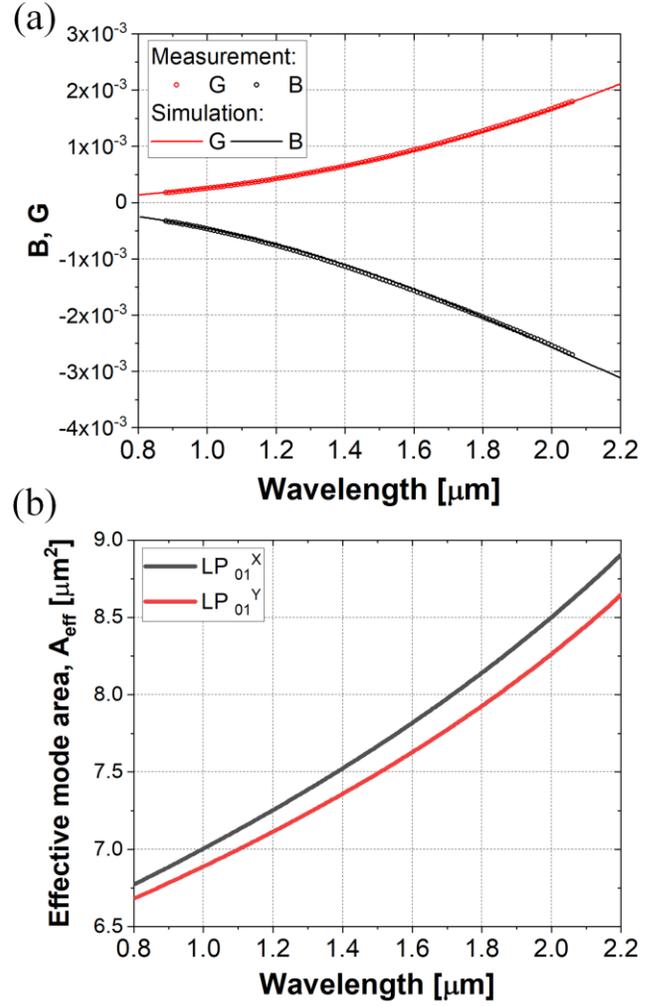

Fig. 5. (a) Measured (dotted curve) and calculated (solid curve) group and phase modal birefringence of the fiber, (b) calculated effective mode area for both orthogonally polarized modes.

To numerically characterize the fiber, we carried out simulations based on the fabricated fibers' geometries. We applied image processing procedures (thresholding and binarization) to the SEM images of the fibers' cross-sections. In the FEM calculation, a mesh composed of 400k triangular elements was used to reproduce the actual geometry of the studied fiber. Additionally, the numerical simulation allows us to calculate the spectral dependence of the effective mode area of both polarization modes used in the numerical simulation of nonlinear fiber response (Fig. 5(b)). The measured values of chromatic dispersion, phase, and group modal birefringence are in excellent agreement with the numerical simulation results (see Figs. 4 and 5(a)). According to the numerical simulations, the fiber supports two spatial modes $LP_{01}$ and $LP_{11}$ in the considered spectral range (up to 1950 nm). However, the presence of the $LP_{11}$ mode does not influence the soliton shift in the fundamental mode since the overlap between the fundamental mode and the $LP_{11}$ modes zeros [24]. The electric field distribution in the two modes have different symmetries: the mode profile is even for the $LP_{01}$ and odd for $LP_{11}$.

## III. SOLITON SHIFT GENERATION AND CHARACTERIZATION RESULTS

### A. Wavelength tunability and coherence

The experimental setup for SSFS generation is depicted in Fig. 6(a). As pump sources, we used an YDFL (Orange-HP, Menlo Systems) delivering 60 fs pulses centered at 1.04 µm with 125 MHz repetition rate, and an in-house built EDFL delivering 25 fs pulses at 1.55 µm (described in detail in Ref. [25]) with 45 MHz repetition rate. The pump beam was coupled to the fiber's core via an aspheric lens (AL) with a 3.1 mm focal length. We used a half-wave plate (λ/2) and a polarization beamsplitter (PBS) for tuning the power coupled into the fiber. Another λ/2 plate was used for alignment of the polarization launched to the fiber. Bandpass filters (F) were used at the output of the fiber to cut-out the residual pump light and transmit only the shifted solitons.

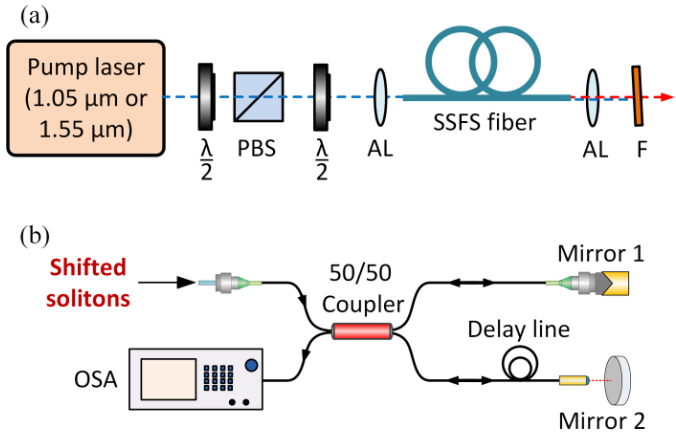

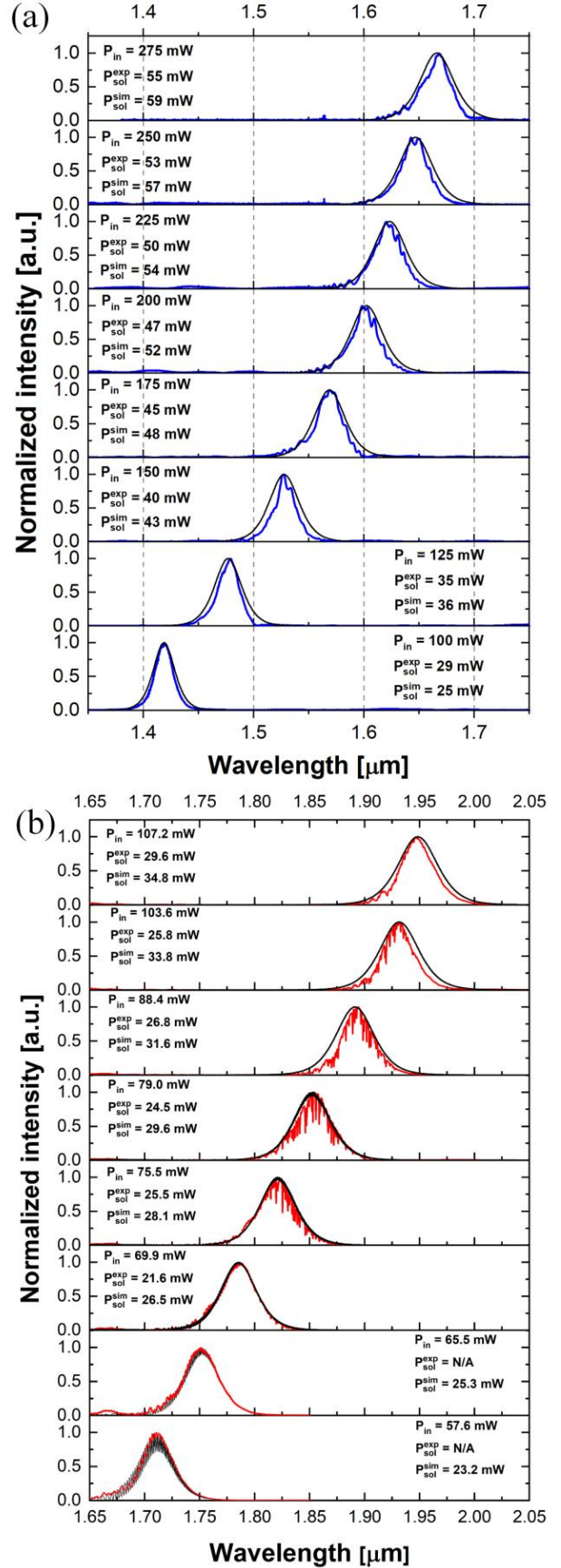

Fig. 6. (a) Experimental setup for SSFS generation; (b) setup of the unequal-path Michelson interferometer used for pulse-to-pulse coherence measurement.

Examples of optical spectra of the shifted solitons generated in the 1.5-m long fiber pumped at 1.04 µm and 1.55 µm are depicted in Figs. 7(a) and 7(b), respectively. All measurements were performed with the input polarization aligned to the slow axis of the fiber. The spectra shown in Fig. 7(a) were recorded using a home-built Fourier-transform spectrometer, while these shown in Fig. 7(b) were recorded with an OSA (Yokogawa AQ6375). The experimental spectra are plotted together with the numerically simulated ones (see below). The observable dips in the 1.8–1.9 µm range in Fig. 7 (b) are the result of water vapor absorption. The center wavelength of the soliton was tuned by only changing the power coupled into the fiber, as indicated in the figures. In the case of 1.04 µm pumping, the input power was changed from 100 to 275 mW. The average power stored in the shifted soliton (filtered out using a long-pass filter with 1350 nm cut-off wavelength, Thorlabs FELH1350) varied between 29 mW (at the shortest wavelengths) and 55 mW (longest wavelengths). The calculated conversion efficiency varied between 20–28%. For the 1.55 µm pumping, the minimum input power needed to obtain the soliton shift was 56 mW, and the maximum available power was 107.2 mW.

Fig. 7. Dynamics of the soliton shift in the 1.5-m-long SSFS fiber under pumping with (a) the 1040 nm YDFL, and (b) the 1560 nm EDFL. Black curves represent the numerically simulated spectra.

The average optical power preserved in the solitons was determined using a filter with the 1750 – 2250 nm passband (Thorlabs FB2000-500). The results indicated the average power of 21.6 mW for the soliton centered at 1786 nm and 29.6 mW for the maximum wavelength shift of 1948 nm. The pump-to-soliton power conversion efficiency estimated for the 1.55 µm pumping was at the level of 27–33%. We emphasize that all efficiency values are calculated with respect to the power of the lasers measured in front of the fiber. Therefore, they include the losses induced by the non-ideal coupling of light to the fiber (which are between 47% and 58%). The conversion efficiency results are comparable to those presented in the literature (22% in. [26], 35% in [11], 35% in [16], 10% in [27]). It has been reported that the efficiency can be maximized even up to 97% [2] when the solitons are generated in an active fiber.

The numerical simulations were performed with a scalar nonlinear Schrödinger solver [28]. The solver was modified to account for the dispersion of the effective mode area [29], Raman response function given by Lin et al. [30], and wavelength-dependent attenuation [31]. We also verified that the scalar approach gives virtually the same results as the vector approach accounting for a vector Raman response [32]. For both pumping sources, we performed simulations of nonlinear pulse propagation over a distance of 1.5 m, observing soliton frequency shifting (Fig. 7). The input powers in simulations correspond to the experimental input powers corrected for the experimentally verified coupling efficiency. In the simulations, we adjusted coupling efficiencies in the ±2% range to fit the solitons spectral positions. Finally, the coupling efficiencies were at the level of approx. 47% and 58% for the EDFL and YDFL, respectively. For 1.55 µm pumping, the initial time profile of the laser pulse was assumed to be the sum of three hyperbolic secant pulses described in [30] to reflect the measured autocorrelation trace. For 1.04 µm pumping, it was assumed that the initial pulse is a Gaussian pulse with full-width at half-maximum (FWHM) of 60 fs. Finally, by integrating the spectral energy density, we obtained the average power carried by soliton pulse. The simulated soliton power $P_{sol}^{sim}$, the experimentally estimated soliton power $P_{sol}^{exp}$ and the measured pumping laser power $P_{in}$ are given in Fig. 7 for each case (for the two lowest laser powers for 1.55 µm pumping, the positions of solitons were outside the passband of the filter and $P_{sol}^{exp}$ were not available). The results are in excellent agreement for both pumping sources. Figure 8 summarizes the maximum achievable shifts (center wavelength of the maximally shifted soliton), power stored in the soliton, and conversion efficiency for both pumping sources.

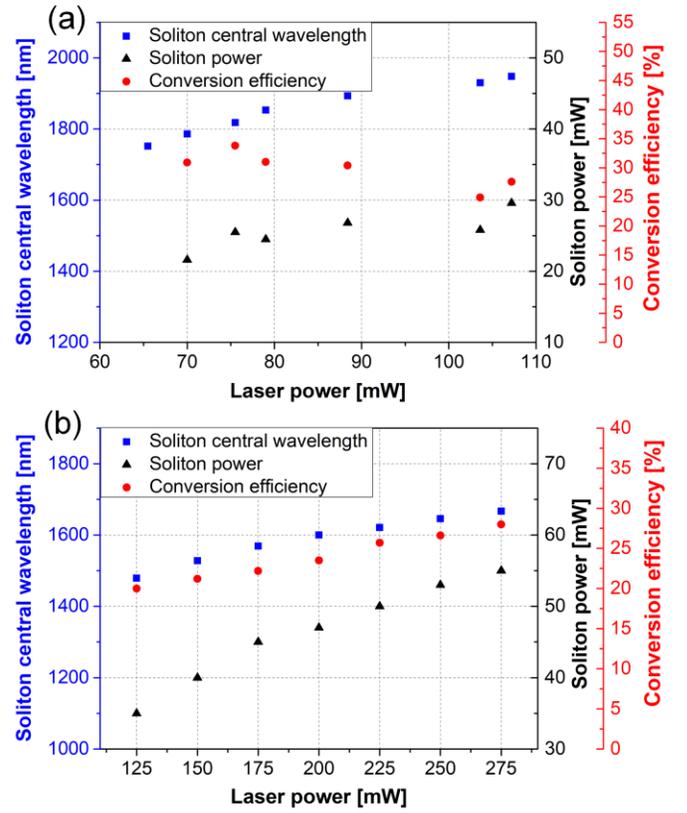

Fig. 8. The soliton wavelength (left scale, blue squares), soliton power (first right scale, black triangles), and the conversion efficiency (second right scale, red dots) as a function of the laser power, measured for (a) 1.04 µm and (b) 1.55 µm pumping wavelength.

The coherence of the generated solitons was investigated using unequal-path fiber-based Michelson interferometers [33], as shown in Fig. 6(b). The pulses were split into two branches using a 50/50% coupler. One arm is longer by $L_{osc}/2$, where $L_{osc}$ is the length of the seed laser cavity. Thus, interference occurs between two consecutive pulses of the pulse train if the phase between the pulses is preserved. This method was previously used for the coherence measurement of solitons [18] supercontinuum sources [33-35] as well as oscillators [36]. The degree of coherence is proportional to the fringe visibility function, defined as $V(\lambda) = [I_{max}(\lambda)-I_{min}(\lambda)]/[I_{max}(\lambda)+I_{min}(\lambda)]$, where $I_{max}$ and $I_{min}$ are the maximum and minimum intensities in the signal, respectively. Both $I_{max}$ and $I_{min}$ used for calculating $V(\lambda)$ were determined by calculating upper and lower envelopes of the spectral interferograms. Figure 9 shows the measured pulse-to-pulse spectral interference patterns at different wavelengths, together with the calculated fringe visibility. The measurements confirm the high coherence of the solitons pumped at 1.04 µm, with $V(\lambda)$ at the level of > 0.7 in the range of 1400 to 1700 nm. In the case of 1.55 µm pumping, the $V(\lambda)$ reaches up to 0.9 for the 1950 nm soliton. We observed a slight degradation of the coherence in the 1.77–1.87 µm range, which is not fully understood and requires further analysis. Nevertheless, the fringes are clearly visible and indicate a coherent frequency shift process.

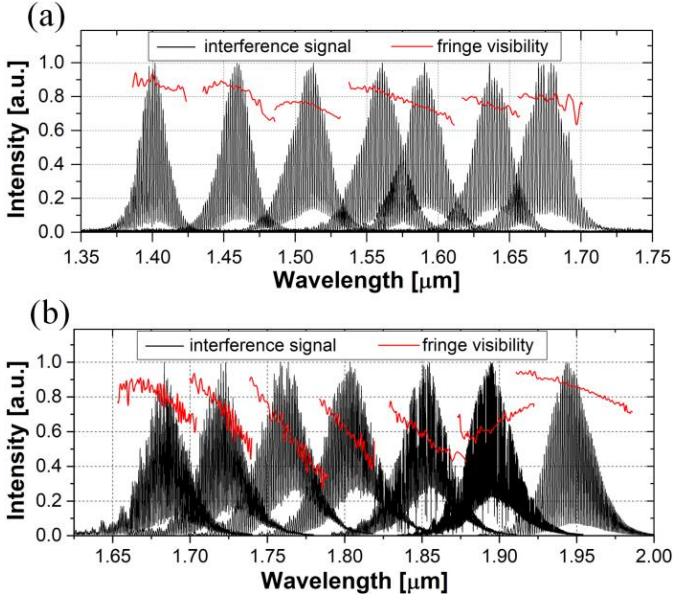

Fig. 9. Measured pulse-to-pulse interference and calculated fringes visibility (red line) of the shifted solitons under pumping with the 1040 nm YDFL (a), and 1560 nm EDFL (b).

Finally, we verified the maximum possible frequency shift for longer fiber samples using the 1.55 μm pump. We tested fiber segments with 180, 210, 230, and 275 cm lengths and recorded the output spectra at the highest available pumping power (107.2 mW). Simultaneously, we verified the coherence of the generated solitons as a function of the fiber length. The results are summarized in Fig. 10. Longer fiber segments enable a larger shift, e.g., up to 2070 nm for a 275 cm long fiber. The shape of the soliton spectra agrees with numerical simulations, shown in Fig. 10(a) (dashed line). In the experiment, we observed degradation of the fringe visibility with increasing fiber length (see Fig. 10(b)). For a 150 cm long segment, V(λ) reaches an average value of 0.9, while for a 275 cm-long fiber, it drops to < 0.5. The degradation of coherence for longer fiber samples might be attributed to the increased impact of modulation instability for increased nonlinear propagation, as shown in [37,38]. The results presented above indicate that the tuning range can be easily increased by using a longer segment of the fiber if necessary. However, as our measurements revealed, there is a trade-off between the broader wavelength range of a soliton shift and the coherence of the shifted soliton. A different – more difficult – approach could be increase the pump power. According to numerical simulations, one needs 136 mW of pumping power (assuming the same coupling efficiency and the pulse profile) to reach 2070 nm in the 150 cm long fiber piece. The same shift was achieved experimentally in 275 cm long fiber with 107.2 mW of pumping power. A longer fiber piece and/or a more powerful laser should allow reaching even 2.4 μm [39]. Further investigation on this subject could be concentrated on the problem of preserving the high coherence level of the widely tuned solitons.

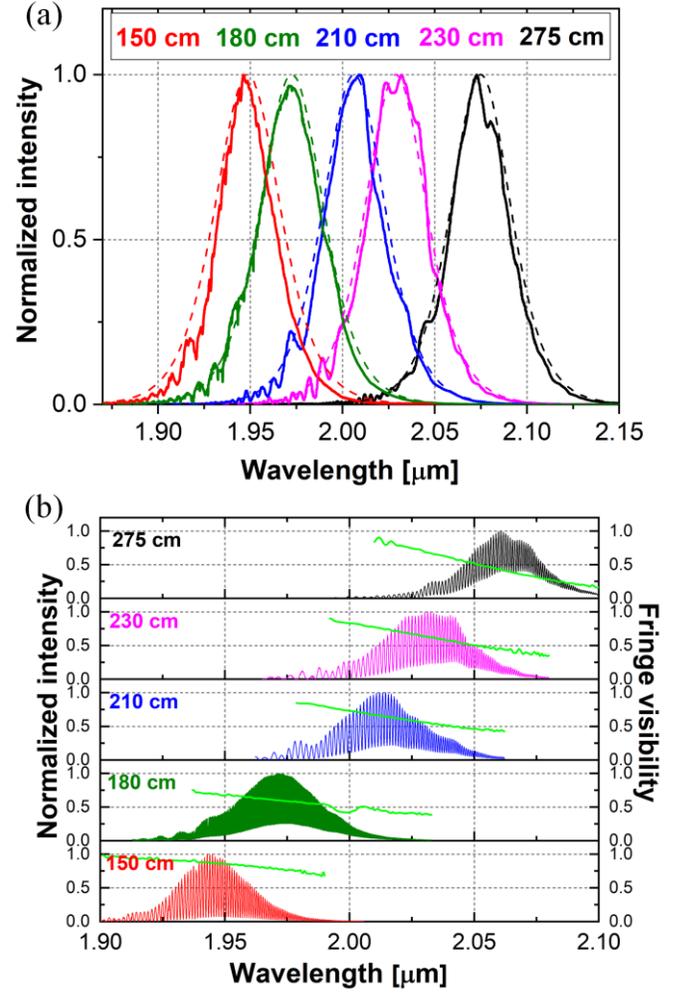

Fig. 10. Wavelength shift vs. SSFS fiber length under 1550 nm pumping: spectra recorded at maximum pumping level for five different lengths (solid lines) together with the numerical simulation (dashed line) (a); pulse-to-pulse coherence measurement for different fiber lengths (b).

### B. Polarization extinction ratio analysis

The polarization extinction ratio (PER) of the solitons was investigated in an experimental setup as depicted in Fig. 11. The half-wave plate (λ/2) placed in front of the fiber was used to match the input polarization with the proper axis of the fiber (slow or fast). The angle of 0° corresponds to the slow axis excitation, while rotation of the plate by 45° excites the fast axis. The polarizer (P) at the output can be aligned either to transmit only light along the slow (0°) or fast axis (90°). The spectrum recorded without the polarizer contains components originating from both polarization axes. The spectra are recorded by the OSA.

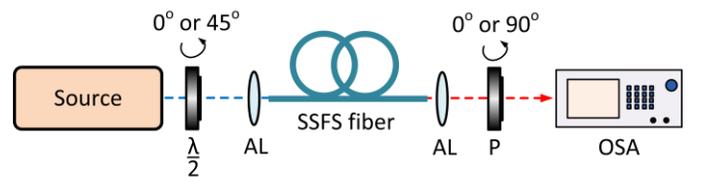

Fig. 11. Experimental setup for PER measurement in the SSFS fiber.

Figure 12 shows the spectra recorded for YDFL pumping at three different power levels, to illustrate the spectral dependence of the soliton's PER. For the input polarization aligned to excite the slow axis of the fiber (plots in the left column, i.e., a, c, e), the PER increases at longer wavelengths, reaching 14 dB for the soliton localized at 1640 nm. The PER in the fast axis (plots in the right column: b, d, f) is lower by 1-2 dB in each case.

trapping effect in the fast polarization axis. This can be explained if we note that in a highly birefringent MOF the group velocity of the slow mode is higher that the group velocity of the fast mode (the opposite of what occurs in conventional fibers). It is also worth mentioning that the spectral position of the soliton and the small spectral component satisfy the group-velocity matching (the group velocity of the small pulse is almost the same as that of the soliton pulse).

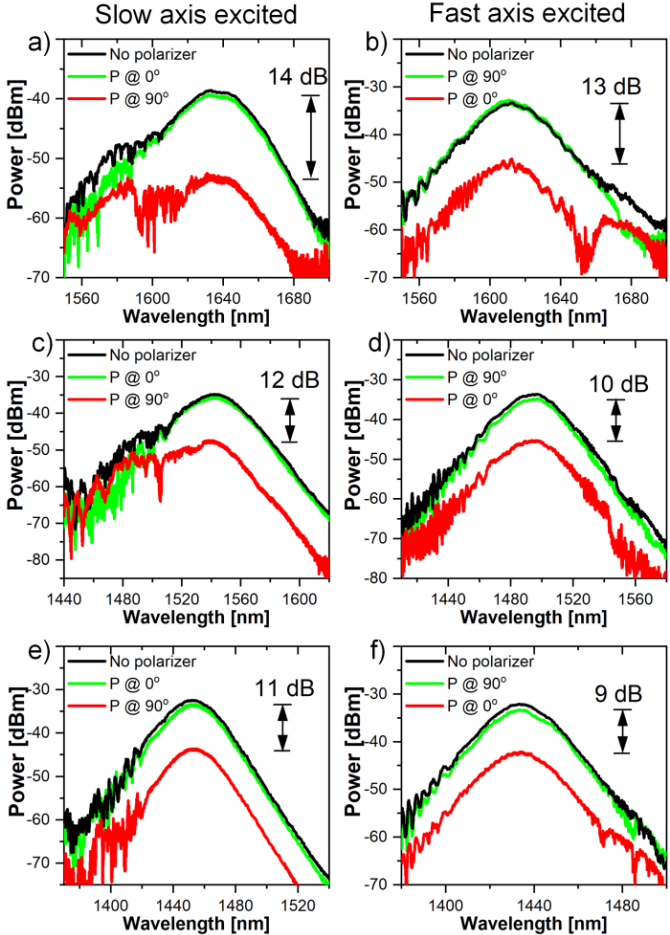

Fig. 12. Soliton PER measurement under YDFL pumping at three different settings of the solitons central wavelength: 1640 nm (a,b), 1540 (c,d), and 1460 (e,f). Plots in the left column are recorded for slow axis excitation, while the right column shows the spectra at the fast axis. Green and red lines represent the spectrum recorded with the output polarizer aligned for maximum and minimum transmission, respectively (the given angles are with respect to the slow axis of the fiber). Black line represents the measurement without the polarizer.

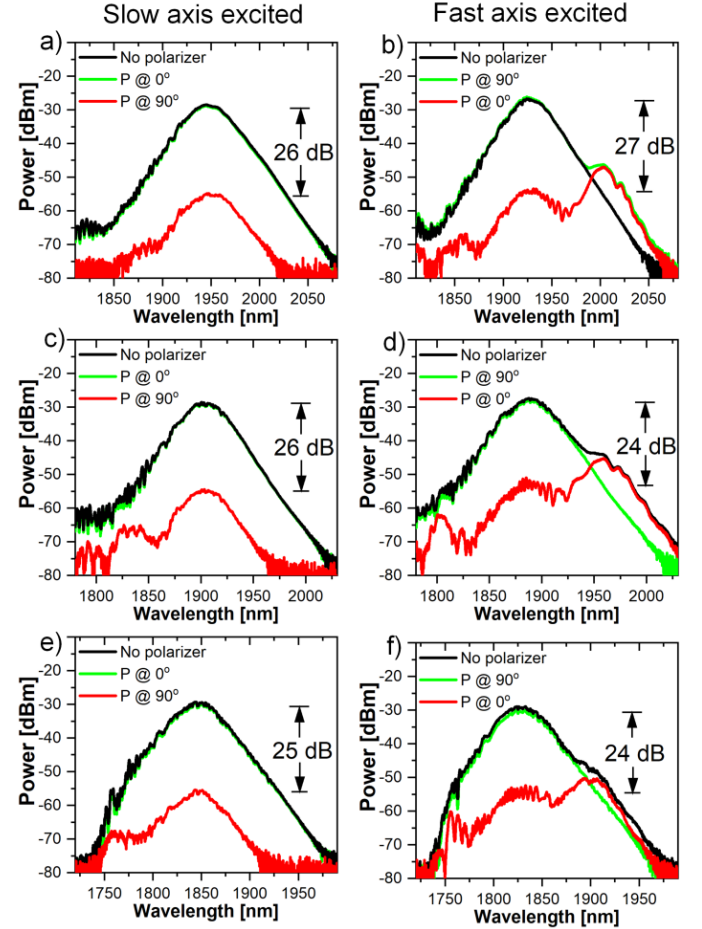

Fig. 13. Soliton PER measurement under EDFL pumping at three different settings of the solitons central wavelength: 1950 nm (a,b), 1900 (c,d), and 1850 (e,f). Plots in the left column are recorded for slow axis excitation, while the right column shows the spectra at the fast axis. Green and red lines represent the spectrum recorded with the output polarizer aligned for maximum and minimum transmission, respectively (the given angles are with respect to the slow axis of the fiber). Black line represents the measurement without the polarizer.

The PER measurement results for EDFL pumping are depicted in Fig. 13. For the polarization direction aligned along the slow axis, the PER always exceeds 25 dB. When the polarization direction is aligned along the fast axis, an additional spectral component is observed at longer wavelength side of the soliton. When the polarizer is aligned to attenuate the fast axis components, the spectral component of the soliton pulse is almost attenuated and PER exceeds 20 dB (Fig. 13 b, d, f). Such a phenomenon of trapping of an orthogonally polarized by the soliton and its shifting towards the longer wavelength side was observed in [5]. However, in contrary to [5], we observed the

## IV. SUMMARY

Summarizing, we have demonstrated a novel type of highly birefringent, microstructured all-silica-based fiber for broadband soliton self-frequency shifting, suitable for pumping at 1.04 μm and 1.55 μm wavelengths. Depending on the used pump source, the input spectrum can be continuously tuned up to 1.67 μm (pump at 1.04 μm) or to 1.95 μm (pump at 1.55 μm) in the same 1.5-m-long fiber sample. Further elongation of the fiber enables longer wavelength shifts (exceeding 2.05 μm). The available tuning range with 1.04 μm pumping covers the wavelength of 1.3 μm, which is needed, e.g., in OCT imaging, and 1.55 μm, which covers the gain bandwidth of Er-doped media. In the case of EDFL pumping, the solitons could cover the range of 1.65 to 1.95 μm, reaching the spectral band covered by Tm-doped gain media. Thus, such a source can act as a seed for Tm-doped fiber amplifiers. The results show that proper dispersion management enables the fabrication of one universal fiber, which will support efficient spectral conversion regardless of the used pump source. For both pumping sources, the conversion efficiency from the fundamental pump pulse to the soliton is higher than 20%. Moreover, the solitons are characterized by a high pulse-to-pulse phase coherence, verified by measuring the fringe visibility of the interference signal of two consecutive signal pulses. In combination with a high polarization extinction ratio, the fiber can be considered as a building block for widely tunable or swept laser sources for application in imaging and spectroscopy.

REFERENCES

[1] F. M. Mitschke and L. F. Mollenauer, "Discovery of the soliton self-frequency shift," *Opt. Lett.*, vol. 11, no 10, pp. 659–661, 1986.
[2] J. Luo, B. Sun, J. Ji, E. Leong Tan, Y. Zhang, and X. Yu, "High-efficiency femtosecond Raman soliton generation with a tunable wavelength beyond 2 μm," *Opt. Lett.*, vol. 42, no. 8, pp. 1568-1571, 2017.
[3] N. G. Horton, K. Wang, D. Kobat, C. G. Clark, F. W. Wise , C. B. Schaffer, and C. Xu, "In vivo three-photon microscopy of subcortical structures within an intact mouse brain," *Nat. Photon.*, vol. 7, pp. 205-209, 2013.
[4] K. F. Lee, C. J. Hensley, P. G. Schunemann, and M. E. Fermann, "Midinfrared frequency comb by difference frequency of erbium and thulium fiber lasers in orientation-patterned gallium phosphide," *Opt. Express*, vol. 25, no. 15, pp. 17411–17416, 2017.
[5] N. Nishizawa and T. Goto, "Widely wavelength-tunable ultrashort pulse generation using polarization maintaining optical fibers," *IEEE J. Sel. Top. Quantum Electron.*, vol. 7, no. 4, pp. 518–524, 2001.
[6] T. Nguyen, K. Kieu, D. Churin, T. Ota, M. Miyawaki, and N. Peyghambarian, "High Power Soliton Self-Frequency Shift With Improved Flatness Ranging From 1.6 to 1.78 μm," *IEEE Photon. Technol. Lett.,* vol. 25, no. 19, 1893-1896, 2013.
[7] F.Tan, H. Shi, R. Sun, P. Wang, and P. Wang, "1 μJ, sub-300 fs pulse generation from a compact thulium-doped chirped pulse amplifier seeded by Raman shifted erbium-doped fiber laser," *Opt. Express*, vol. 24, no. 20, pp. 22461-22468, 2016.
[8] H. Delahaye, G. Granger, M. Jossent, J. Gomes, L. Lavoute, D. Gaponov, and S. Février, "Nanojoule sub-100 fs Mid Infrared Pulse Generated From a Fully Fusion-Spliced Fiber Laser," in *Advanced Photonics 2018 (BGPP, IPR, NP, NOMA, Sensors, Networks, SPPCom, SOF)*, OSA Techncal Digest (online) (Optical Society of America, 2018), paper NpM4C.6.
[9] M. Yu. Koptev, E. A. Anashkina, A. V. Andrianov, V. V. Dorofeev, A. F. Kosolapov, S. V. Muravyev, and A. V Kim, "Widely tunable mid-infrared fiber laser source based on soliton self-frequency shift in microstructured tellurite fiber," *Opt. Lett.*, vol. 40, no. 17, pp. 4094–4097, 2015.
[10] E.A. Anashkina, M.Y. Koptev, S.V. Muravyev, V.V. Dorofeev, A.V. Andrianov, and A.V. Kim, "Raman soliton generation in microstructured tellurite fiber pumped by hybrid Erbium/Thulium fiber laser system," *J. Phys.: Conf. Ser.*, vol. 735, 012020, 2016.
[11] L. Tang, G. Wright, K. Charan, T. Wang, C. Xu, and F. W. Wise, "Generation of intense 100 fs solitons tunable from 2 to 4.3 μm in fluoride fiber," *Optica*, vol. 3, pp. 948–951, 2016.
[12] H. Lim, J. Buckley, A. Chong, and F.W. Wise, "Fibre-based source of femtosecond pulses tunable from 1.0 to 1.3 μm," *Electron. Lett.*, vol. 40, no. 24, pp. 1523–1525, 2004.
[13] J. Dawlaty, A. Chong, and F. Rana, "Widely Tunable (1.0-1.7 μm) Yb-doped Fiber Mode-Locked Laser Source with ˜100 fs Pulse Widths based on Raman Frequency Shift," 2007 Digest of the IEEE/LEOS Summer Topical Meetings, pp. 157–158, 2007.
[14] J. Takayanagi, T. Sugiura, M. Yoshida, and N. Nishizawa, "1.0–1.7-μm Wavelength-Tunable Ultrashort-Pulse Generation Using Femtosecond Yb-Doped Fiber Laser and Photonic Crystal Fiber," *IEEE Photon. Technol. Lett.*, vol. 18, no. 21, pp. 2284-2286, 2006.
[15] J. Jiang, A. Rueh, I. Hartl, and M. E. Fermann, "Tunable coherent Raman soliton generation with a Tm-fiber system," in *CLEO:2011 - Laser Applications to Photonic Applications*, 2011, paper CThBB5.
[16] T. W. Neely, T. A. Johnson, and S. A. Diddams, "High-power broadband laser source tunable from 3.0 μm to 4.4 μm based on a femtosecond Yb:fiber oscillator," *Opt. Lett.*, vol. 36, no. 20, pp. 4020–4022, 2011.
[17] D. Deng, T. Cheng, X. Xue, H. Tuan Tong, T. Suzuki, Y. Ohishi, "Widely tunable soliton self-frequency shift and dispersive wave generation in a highly nonlinear fiber," in *Proc. SPIE 9359, Opt. Components and Materials XII*, 2015, paper 935903.
[18] G. Soboń, T. Martynkien, P. Mergo, L. Rutkowski, and A. Foltynowicz, "High-power frequency comb source tunable from 2.7 to 4.2 μm based on difference frequency generation pumped by an Yb-doped fiber laser," *Opt. Lett.*, vol. 42, no. 9, pp. 1748–1751, 2017.
[19] K. Suzuki, H. Kubota, S. Kawanishi, M. Tanaka, and M. Fujita, "Optical properties of a low-loss polarization-maintaining photonic crystal fiber," *Opt. Express,* vol. 9, no. 13, pp. 676–680, 2001.
[20] I. Gris-Sánchez, B.J. Mangan, and J.C. Knight, "Reducing spectral attenuation in small-core photonic crystal fibers," Opt. Mater. Express, vol. 1, no.2, pp.179-184, 2011.
[21] P. Hlubina, M. Kadulova, P. Mergo, "Chromatic dispersion measurement of holey fibres using a supercontinuum source and a dispersion balanced interferometer," *Opt. Lasers Eng.*, vol. 51, no. 4, pp. 421–425, 2013.
[22] P. Hlubina and D. Ciprian, "Spectral-domain measurement of phase modal birefringence in polarization-maintaining fiber," *Opt. Express*, vol. 15, no. 25, pp. 17019–17024, 2007.
[23] P. Hlubina, D. Ciprian, and M. Kadulova, "Wide spectral range measurement of modal birefringence in polarization-maintaining fibres," *Meas. Sci. Technol.*, vol. 20, no. 2, 025301, 2009.
[24] P. Horak and F. Poletti, "Multimode nonlinear fibre optics: theory and applications," in Recent Progress in Optical Fiber Research, M. Yasin, S. W. Harun, and H. Arof, eds. (IntechOpen, 2012), Chap. 1
[25] J. Sotor and G. Soboń, "24 fs and 3 nJ pulse generation from a simple, all polarization maintaining Er-doped fiber laser," *Laser Phys. Lett.*, vol. 13, no. 12, 125102, 2016.
[26] S. A. Dekker, A. C. Judge, Ravi Pant, Itandehui Gris-Sánchez, Jonathan C. Knight, C. Martjn de Sterke, and Benjamin J. Eggleton, "Highly-efficient, octave spanning soliton self-frequency shift using a specialized photonic crystal fiber with low OH loss," *Opt. Express,* vol. 19, no. 18, pp. 17766-17773, 2011.
[27] G. Soboń, T. Martynkien, D. Tomaszewska, K. Tarnowski, P. Mergo, and J. Sotor, "All-in-fiber amplification and compression of coherent frequency-shifted solitons tunable in the 1800–2000 nm range," *Photon. Res.,* vol. 6, no. 5, pp. 368-372, 2018.
[28] J.C. Travers, M.H. Frosz, J.M. Dudley, "Nonlinear fibre optics overview," in *Supercontinuum Generation in Optical Fibers*, Cambridge: Cambridge University Press, 2010, pp. 32-51.
[29] J. Laegsgaard, "Mode profile dispersion in the generalised nonlinear Schrödinger equation", *Opt. Express,* vol. 15, no. 24, pp. 16110–16123, 2007.
[30] Q. Lin and G. P. Agrawal, "Raman response function for silica fibers," *Opt. Lett.*, vol. 31, no. 21, pp. 3086–3088, 2006.
[31] O. Humbach, H. Fabian, U. Grzesik, U. Haken, W. Heitmann, "Analysis of OH absorption bands in synthetic silica," *J. Non-Cryst. Solids,* vol. 203, pp. 19–26, 1996.
[32] P. Balla and G. P. Agrawal, "Nonlinear interaction of vector solitons inside birefringent optical fibers," *Phys. Rev. A,* vol. 98, no. 2, 023822, 2018.
[33] J. W. Nicholson and M. F. Yan, "Cross-coherence measurements of supercontinua generated in highly-nonlinear, dispersion shifted fiber at 1550 nm," *Opt. Express,* vol. 12, no. 4, pp. 679–688, 2004.
[34] K. Tarnowski, T. Martynkien, P. Mergo, J. Sotor, and G. Soboń, "Compact all-fiber source of coherent linearly polarized octave-spanning supercontinuum based on normal dispersion silica fiber," *Sci. Rep.,* vol. 9, 12313, 2019.
[35] S. Kim, J. Park, S. Han, Y.-J. Kim, and S.-W. Kim, "Coherent supercontinuum generation using Er-doped fiber laser of hybrid mode-locking," *Opt. Lett.*, vol. 39, no. 10, pp. 2986–2989, 2014.

[36] A. F. J. Runge, C. Aguergaray, N. G. R. Broderick, and M. Erkintalo, "Coherence and shot-to-shot spectral fluctuations in noise-like ultrafast fiber lasers," *Opt. Lett.*, vol. 38, no. 10, pp. 4327–4330, 2013.

[37] J. Dudley and S. Coen, "Numerical simulations and coherence properties of supercontinuum generation in photonic crystal and tapered optical fibers," *IEEE J. Sel. Top. Quantum Electron.*, vol. 8, no. 3, pp. 651–659, 2002.

[38] G. Soboń, R. Lindberg, V. Pasiskevicius, T. Martynkien, and J. Sotor, "Shot-to-shot performance analysis of an all-fiber supercontinuum source pumped at 2000 nm," *J. Opt. Soc. Am. B,* vol. 36, no. 2, pp. A15-A21, 2019.

[39] B. Li, M. Wang, K. Charan, M. Li, and C. Xu, "Investigation of the long wavelength limit of soliton self-frequency shift in a silica fiber," Opt. Express, vol. 26, no. 15, pp. 19637-19647, 2018